\documentclass[conference,compsoc]{IEEEtran}

\IEEEoverridecommandlockouts
\usepackage[backend=biber,style=ieee]{biblatex}
\usepackage{siunitx}
\usepackage{multirow}
\usepackage{amsmath,amssymb,amsfonts}
\usepackage{algorithm}
\usepackage{algpseudocode}
\usepackage{graphicx}
\usepackage{textcomp}
\usepackage{xcolor}
\usepackage{url}
\usepackage{amsthm}
\usepackage{subcaption}
\theoremstyle{definition}

\def\BibTeX{{\rm B\kern-.05em{\sc i\kern-.025em b}\kern-.08em
    T\kern-.1667em\lower.7ex\hbox{E}\kern-.125emX}}

\begin{document}

\title{Model-Based Calculation Method of Mining Fairness in Blockchain}

\author{\IEEEauthorblockN{Akira Sakurai}
\IEEEauthorblockA{\textit{Kyoto University} \\
Kyoto, Japan}
\and
\IEEEauthorblockN{Kazuyuki Shudo}
\IEEEauthorblockA{\textit{Kyoto University} \\
Kyoto, Japan}
}

\maketitle

\begin{abstract}
Mining fairness in blockchain refers to equality between the computational resources invested in mining and the block rewards received. There exists a dilemma wherein increasing the transaction processing capacity of a blockchain compromises mining fairness, thereby undermining its decentralization. This dilemma remains unresolved despite methods such as the greedy heaviest observed subtree (GHOST) protocol, indicating that mining fairness is an inherent bottleneck in the transaction processing capacity of the blockchain system. However, despite its significance, existing analyses neglect the impact of blockchain forks, resulting in imprecise evaluations and limited insights. To address this issue, we propose a method for calculating mining fairness that explicitly captures the influence of forks. First, we approximate a complex blockchain network using a simple mathematical model, assuming that no more than two blocks are generated per round. Within this model, we quantitatively determine local mining fairness and derive several measures of global mining fairness based on local mining fairness. Subsequently, we validated by blockchain network simulations that our calculation method computes mining fairness in networks much more accurately than existing methods. The proposed method facilitates a rigorous evaluation of trade-offs between scalability and decentralization by offering a clear, quantitative framework for measuring and comparing reward distribution among miners. Consequently, it is expected to provide valuable insights for future mining fairness research and the design of next-generation blockchain systems.
\end{abstract}

\IEEEpeerreviewmaketitle

\section{Introduction}
Blockchain is a foundational technology primarily used in decentralized currency systems such as Bitcoin~\cite{bitcoin}. In blockchain systems, transactions are processed in units known as blocks. Generating a block involves numerous hash calculations, a process referred to as mining. Nodes that perform mining are called miners. Each miner follows a fork choice rule to identify and extend the main chain. When miners successfully generate a block, they may be rewarded via a coinbase transaction, by which they would receive what is known as a block reward. However, these block rewards are obtainable only when the blocks that have been generated become part of the main chain.

Mining fairness refers to equality between the computational resources invested in mining and the resulting block rewards; that is, it is equality between the proportion of hashrate and the proportion of block rewards (hereafter referred to as the block reward rate). If all blocks were incorporated into the main chain, mining fairness would be achieved because the number of blocks generated by each miner would not be affected by the state of the network. However, in practice, not all blocks are included in the main chain because of blockchain forks, and mining fairness is compromised when blocks are discarded. Forks can be classified into two types: intentional (malicious) and unintentional. The latter occurs when multiple blocks are generated almost simultaneously. This study addresses mining fairness in the context of unintentional forks.

\begin{figure}[tb]
\begin{center}
\includegraphics[width= \linewidth]{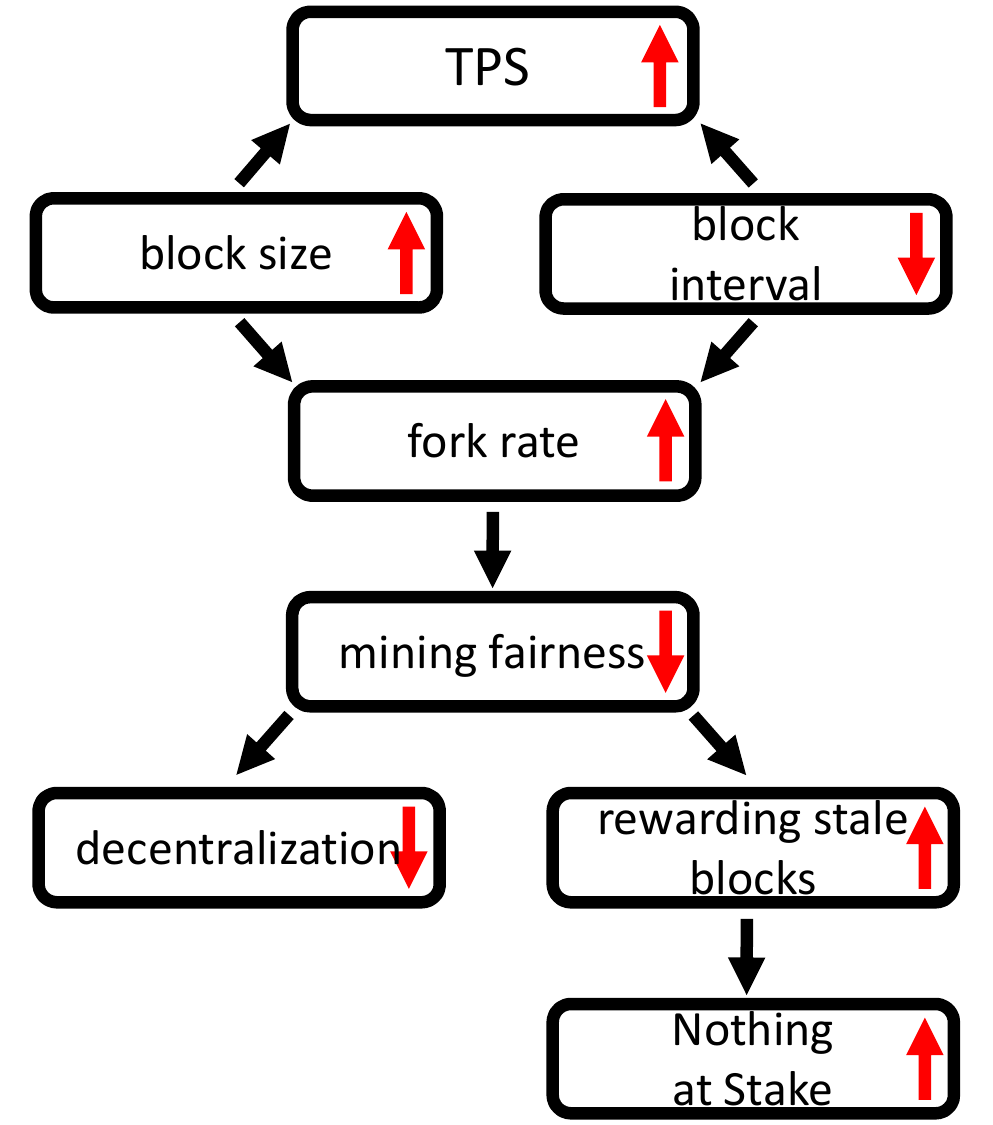}
\end{center}
\caption{Schematic of mining fairness, illustrating how it establishes a link between transaction processing capacity (TPS) and decentralization. Even if mining fairness is enhanced by rewarding stale blocks, the nothing-at-stake problem emerges, compromising security.}
\label{miningfairnessisimportant}
\end{figure}

Mining fairness introduces a trade-off between the transaction processing capacity and decentralization in blockchain systems (Fig.~\ref{miningfairnessisimportant})—increasing the transaction processing capacity of a system compromises decentralization. 
The transaction processing capacity depends on the number of transactions processed per block and the block generation interval. To increase this capacity, one might increase the block size and reduce the block generation interval. However, it is well known that increasing block sizes and reducing block generation intervals result in higher fork rates~\cite{informationpropagation}. As observed previously, an increase in the fork rate undermines mining fairness. If mining fairness is reduced, some miner groups achieve higher profit rates than others. Consequently, miners with lower profit rates end up leaving the system, whereas those with higher profit rates expand, leading to centralization and reduced decentralization.

The dilemma between transaction processing capacity and decentralization in blockchain systems that arises from mining fairness has yet to be resolved. In this context, mining fairness is an inherent bottleneck in the transaction processing capacity. Here, we demonstrate that the dilemma caused by mining fairness is inherent, using the countermeasures adopted by Ethereum~\cite{ethereum} (modified greedy heaviest observed subtree (GHOST) protocol~\cite{GHOST}) as an example. Increasing the transaction processing capacity of a blockchain leads to more forks, which, in turn, causes two main problems. First, there is an increased risk of attacks such as double-spending attacks and selfish mining~\cite{majorityisnotenough, onTheSecurity}. Second, mining fairness is compromised. To address the first problem, Ethereum has introduced the GHOST protocol. In addition, to address the second problem regarding the impact on mining fairness, Ethereum partially rewards blocks that cause forks but are not incorporated into the main chain (stale blocks). However, this approach has its own challenges. From another perspective, this implies that even blocks that cause forks can receive block rewards, thereby reinforcing economic incentives for attacks such as double-spending attacks and selfish mining. This situation shares the same structure as the nothing-at-stake problem. It is well known that, in Ethereum, the risk of attacks that compromise mining fairness, including selfish mining~\cite{majorityisnotenough}, is increased~\cite{Ritz_2018, UncleBlockMechanismEffectonEthereumSelfishndStubbornMining, UBA}; this indicates that the measures taken by Ethereum do not fundamentally solve the problems related to mining fairness.

Thus, it is crucial to perform further analyses on mining fairness. One possible approach to these analyses is to perform simulations, which, unfortunately, is time-consuming and impractical. Consequently, alternative approaches have been explored~\cite{OnScalingDecentralizedBlockchains, Kanda2020BlockIA, tamingprop, DotheRichGetRicher, ModelingtheImpactofNetwork, ImpactofTemporaryFork, CharacterizingtheImpactofNetworkDelayonBitcoinMining, LessIsMore}. However, these methods do not accurately account for how mining fairness is compromised by forks (Section~\ref{existinganalyses}) and, consequently, lead to analyses based on only a weak reflection of real-world systems, making it challenging to derive meaningful insights from them.

In this study, we propose a model-based calculation method for quantitatively analyzing mining fairness. We approximate a blockchain network using a simplified model, assuming that each round contains at most two blocks. A round $r$ is defined as a unit of time that starts with the generation of a block at height $r$. In other words, we assume that at most one fork occurs per round. By modeling the blockchain network based on our concept of rounds, we can more accurately account for the impact of blockchain forks, thereby enabling a much more precise calculation of mining fairness.

Subsequently, we validate the accuracy of our proposed model-based calculation in measuring mining fairness by conducting simulation experiments. However, validating our method in large-scale networks would be challenging due to computational constraints. Hence, we perform the validation in networks with small numbers of miners. Our results demonstrate that the model-based calculation quantitatively determines mining fairness much more accurately than existing methods.

In Section \ref{roundsection}, we introduce the concept of rounds. It becomes possible to achieve a more accurate calculation of mining fairness by capturing the impact of forks on rounds. Section \ref{existinganalyses} discusses related work. Section \ref{mbcsection} describes the proposed method, the model-based calculation. Section \ref{mbcvalidation} presents the validation of the proposed method. Section \ref{conclusion} provides the conclusion.

\section{Rounds} \label{roundsection}
\begin{figure}[tb]
\begin{center}
\includegraphics[width= \linewidth]{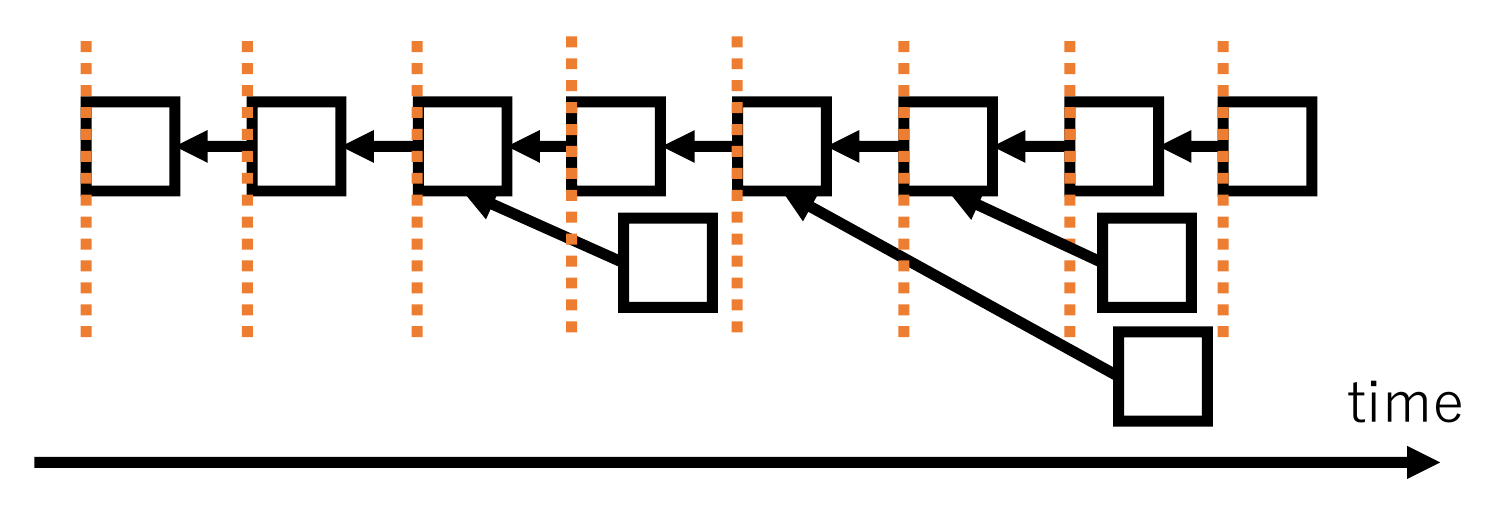}
\end{center}
\caption{Rounds in a blockchain. Each interval enclosed by consecutive orange dotted lines represents a single round in the blockchain.}
\label{defround}
\end{figure}

Herein, we introduce the concept of rounds in a blockchain to accurately capture and incorporate the impact of forks. A round is a time interval defined from a global perspective. Specifically, round $r$ refers to the time from the first generation of a block at height $r$ to the first generation of a block at height $r+1$. The height of round $r$ is defined as $r$.

Because forks occur probabilistically within a blockchain network, the number of blocks per round does not always equal one. After the first block at height $r$ has been generated, another miner may generate a new block before the first block is propagated throughout the network. Notably, not all blocks generated in round $r$ have a height of $r$. For example, a miner unaware of the block at height $r-1$ will generate a block at height $r-1$ during round $r$ (Fig.~\ref{defround}).

The round start rate of each miner is defined as the probability that the miner initiates a round. Note that, owing to the occurrence of forks, the round start rate does not equal the proportion of the hashrate.

With the introduction of rounds, we can formally define the fork rate. The fork rate is the probability that the number of blocks per round is two or more. When the number of blocks per round is two or more, the blockchain diverges or forks. Therefore, forks (and fork rates) are named as such because they function similarly to a traditional fork.

\section{Related Work} \label{existinganalyses}
\subsection{Mining Fairness for Unintentional Forks}
Croman et al.~proposed several metrics and solutions to address the scalability problems of blockchains \cite{OnScalingDecentralizedBlockchains}. They highlighted mining fairness as a potentially more valuable metric despite its difficulty in measurement.

Since then, numerous studies have attempted to analyze mining fairness~\cite{Kanda2020BlockIA, tamingprop, DotheRichGetRicher, 
ModelingtheImpactofNetwork, ImpactofTemporaryFork, CharacterizingtheImpactofNetworkDelayonBitcoinMining, LessIsMore}; however, none have accurately captured the impact of forks on mining fairness. Consequently, the analyses conducted thus far diverged from the actual values observed in blockchain networks. Herein, we discuss each of these studies in detail.

Kanda et al.~introduced the concept of effective hashrate, based on the idea that miners cannot contribute to the main chain until they receive the latest block~\cite{Kanda2020BlockIA}. They calculated the effective hashrate by multiplying the original hashrate by the ratio of the time required to receive a block to the block generation interval. They asserted that the proportion of the effective hashrate equals the block reward rate.
Jiang et al.~analyzed mining fairness based on a concept similar to that proposed by Kanda et al.~\cite{tamingprop}. They calculated the average block reception time for each miner and defined the maximum difference in the average block reception times as mining fairness. However, their concept of mining fairness evidently deviates from reality.

Xiao et al.~proposed a model-based approach to analyzing mining fairness in a blockchain network to analyze the blockchain network connectivity~\cite{ModelingtheImpactofNetwork}. The main difference between their approach and our model-based calculation is that the round start rate is considered in the proposed method. Conversely, they assumed that the round start rate equals the proportion of hashrate, which is not necessarily true.

Chen et al.~examined the impact of forks on the hashrate distribution of blockchain networks from the perspective of mining fairness~\cite{ImpactofTemporaryFork}. Their analysis had two main limitations. First, they did not consider the impact of forks on the round start rate. They assumed that the round start rate equals the proportion of the hashrate, which our analysis has demonstrated to be incorrect. Second, they considered only the block rewards for the miner who initiated the round and did not consider the rewards for subsequent blocks in the round.

Mao et al.~investigated how the manner in which miners are connected affects mining fairness~\cite{LessIsMore}. They conducted both theoretical analyses and simulations; however, each approach had limitations. Regarding their theoretical analysis—they considered only a few blocks directly connected to the genesis block. Conversely, regarding their simulation analysis—they ran a scenario where the ratios of the propagation time to the average block generation interval, $d/T$, were extremely high. Although this condition might allow the observation of trends related to mining fairness, it does not accurately reflect the behavior of actual blockchain networks, raising concerns about the generalizability of the results.

Huang et al.~compared the proof of work (PoW) and proof of stake (PoS) from the perspective of mining fairness~\cite{DotheRichGetRicher}. They claimed that mining fairness is achieved in PoW blockchains in that the block reward rate equals the proportion of the hashrate. However, this conclusion arose because they did not consider unintentional forks in PoW blockchains. They also analyzed the convergence rate in addition to the expected value of mining fairness. Conversely, in this study, we analyzed only the expected value of mining fairness.

\subsection{Mining Fairness for Intentional Forks}
Attacks that intentionally compromise mining fairness to increase block reward rates unjustly have been studied extensively~\cite{rosenfeld2011analysis, majorityisnotenough, FAW, StubbornMining}. For example, Eyal et al.~proposed a mining strategy known as selfish mining, which increases the block reward rate by intentionally causing forks \cite{majorityisnotenough}. They defined the success of selfish mining as achieving a block reward rate that exceeds the proportion of the hashrate, indicating that local mining fairness becomes positive. Sun et al.~introduced the Kullback--Leibler (KL) divergence between the distributions of block reward rates and proportions of hashrate as a measure of the impact of selfish mining~\cite{TFTstrategy}. This concept is similar to global mining fairness. However, while KL divergence measures the difference between distributions, we are interested in the distribution of the difference between block reward rates and the proportions of hashrate. The KL divergence can be zero even when mining fairness is compromised. Therefore, their definition of mining fairness has limited expressiveness. Consequently, this study does not address mining fairness based on how they defined it.

\section{Model-Based Calculation} \label{mbcsection}
We propose a model-based calculation of mining fairness. In this approach, we replace the complex real-world blockchain network with a simpler model in which at most two blocks can be generated per round. Here, a ``round" refers to the concept introduced in Section~\ref{roundsection}. This simplification allows us to appropriately account for the impact of forks on mining fairness, thereby significantly improving the accuracy of mining fairness calculation.

\subsection{Model}
We approximate a complex blockchain network using a simplified model to calculate mining fairness. First, we define the set of miners as $V$, and let $N$ be the number of elements in $V$. The proportion of the hashrate of miner $i$ ($i \in V$) is denoted by $\alpha_i$. When a new block is generated within the network, the probability that miner $i$ has generated that block is equal to the proportion of their hashrate, $\alpha_i$. The number of blocks generated in each round is assumed to be at most two, implying that there is at most one fork per round. We assume that the block rewards are equal.

Let $F_{ij}$ be the probability that miner $j$ generates a block that causes a fork within round $r$ started by miner $i$. After a fork occurs, as additional blocks are generated, one of the blocks will be incorporated into the main chain, while the other will not be. Let $W_{ij}$ be the probability that the block generated by miner $i$ is incorporated into the main chain under the conditions that (a)~round $r$ starts with the block generated by miner $i$ and (b)~miner $j$ generates a block that causes a fork.

\subsection{Definition of Mining Fairness}
Before calculating mining fairness, we first define it. In this study, mining fairness is divided into local mining fairness and global mining fairness. We define local mining fairness, $LF$, based on two measures, as follows:
\begin{align}
LF_1(i) &= r_i - \alpha_i, \\
LF_2(i) &= \frac{LF_1(i)}{\alpha_i},
\end{align}
where $r_i$ refers to the block reward rate for each miner, $LF_1$ denotes the profit of each miner, and $LF_2$ denotes the profit rate of each miner.

Next, we define global mining fairness, $GF$, using local mining fairness, as follows:
\begin{align}
GF_1 &= \sum_{i \in V} LF_1(i) \quad (LF_1(i) > 0), \label{gf1}\\
GF_2 &= \max_{i \in V} LF_2(i) - \min_{i \in V} LF_2(i), \label{gf2}
\end{align}
where $GF_1$ is the sum of the $LF_1$ values that are positive, and $GF_2$ is the maximum difference in the profit rates. Other mining fairness measures can also be defined using $LF$.

Local mining fairness is particularly useful for individual miners, while global mining fairness is important for system designers and engineers. For instance, miners aim to select the most profitable strategies, which is inherently equivalent to improving local mining fairness. On the other hand, system designers seek to establish a fair mining ecosystem, making global mining fairness a crucial objective.

\subsection{Calculation of Mining Fairness} \label{calculating}
This section demonstrates a computational method for mining fairness based on the previously presented model. First, we determine the round start rate. Next, we calculate the local mining fairness $LF_1$, which is the difference between each miner's block reward rate and the proportion of the hashrate. We also determine each miner's profit rate $LF_2$. After the local mining fairness has been calculated, the global mining fairness can be easily determined.

Let $X_r$ be a random variable representing the miner that generates the block that starts the round $r$. Then, the following equation holds:
\begin{align}
&P(X_{r + 1} = i) = \nonumber\\
&\sum_{j \in V} \left(\alpha_i (1 - F_{ji}) + \sum_{k \in V} \alpha_k F_{jk} \alpha_i \right) P(X_{r} = j). \label{repeat}
\end{align}

Notably, $P(X_{r + 1} = i)$ is only dependent on $P(X_{r} = j)$. Therefore, the stochastic process $\{X_r\}_{r = 0}^\infty$ is a Markov chain. Additionally, this Markov chain is ergodic in most cases because $F_{ij}$ is less than 1 and $\alpha_i (1 - F_{ji}) + \sum_{k} \alpha_k F_{jk} \alpha_i$ is usually positive. Consequently, a unique stationary distribution exists, and the limit distribution is stationary. We can then determine the stationary distribution by iterating (\ref{repeat}).

Let the limit distribution be $\pi$. This represents the distribution of the miners that generate blocks that start rounds after sufficient time has passed. Using $\pi$, the block reward rate $r_i$ for each miner is given by the following equation:
\begin{align}
r_i =& \pi(i)(1 - \sum_{j \in V} \alpha_j F_{ij} + \sum_{j \in V} \alpha_j F_{ij} W_{ij}) \nonumber\\
    &+ \sum_{j \in V} \pi(j) \alpha_i F_{ji} (1 - W_{ji}).
\end{align}
Thus, $LF_1$ of miner $i$ is as follows:
\begin{align}
LF_1(i) =& r_i - \alpha_i \\
=& \pi(i)(1 - \sum_{j \in V} \alpha_j F_{ij} + \sum_{j \in V} \alpha_j F_{ij} W_{ij}) \nonumber\\
&+ \sum_{j \in V} \pi(j) \alpha_i F_{ji} (1 - W_{ji}) - \alpha_i, \label{lf1}
\end{align}
whereas $LF_2$ can be calculated as follows:
\begin{align}
LF_2(i) = \frac{LF_1(i)}{\alpha_i}. \label{lf2}
\end{align}

\subsection{Algorithm}
In this section, we describe the algorithm used in this paper study to calculate mining fairness, as detailed in Section \ref{calculating}.

Algorithm \ref{calculating the mining fairness} employs an iterative method to compute the mining fairness for each miner. The round start rate calculation is performed between lines 9 and 27. Specifically, the fork rate is precomputed between lines 10 and 16. The variable \textit{loop} manages the operations executed in each iteration. The calculations within the \textbf{for} loop from lines 20 to 26 follow the same process as described in (\ref{repeat}). Mining fairness is computed between lines 28 and 34, with the \textbf{for} loop calculations corresponding to Equations (\ref{lf1}) and (\ref{lf2}).

\begin{algorithm}
\caption{Calculation of local mining fairness}
\label{calculating the mining fairness}
\begin{algorithmic}[1]
\Statex \textbf{The following variables are provided by the model: }
\State $V$ \Comment{set of miners}
\State $N$ \Comment{number of miners}
\State $\alpha[N]$ \Comment{proportion of hashrate}
\State $F[N][N]$ \Comment{fork rate}
\State $W[N][N]$ \Comment{winning rate}
\Statex
\Statex \textbf{Our goal is to calculate the following values:}
\State $\pi[N][2]$ \Comment{round start rate}
\State $LF_1[N]$ \Comment{$LF_1$}
\State $LF_2[N]$ \Comment{$LF_2$}
\Statex
\Statex \textbf{Calculating the round start rate of each miner:}
\State $\epsilon$ \Comment{error}
\State $dp[N]$ \Comment{for dynamic programming}
\For{$i \in V$}
    \State $dp[i] \gets 0$
    \For{$j \in V$}
        \State $dp[i] \gets dp[i] + \alpha[j] F[i][j]$
    \EndFor
\EndFor
\State $loop \gets 0$
\While{$\exists i \in V$ s.t. $|\pi[i][loop\ \bmod\ 2] - \pi[i][(loop + 1)\ \bmod\ 2]| > \epsilon$}
    \State $loop \gets (loop + 1)\ \bmod\ 2$
    \For{$i \in V$}
        \State $\pi[i][(loop + 1)\ \bmod\ 2] \gets 0$
        \For{$j \in V$}
            \State $\pi[i][(loop + 1)\ \bmod\ 2] \gets \pi[i][(loop + 1)\ \bmod\ 2] + \alpha[i] (1 - F[j][i]) \pi[j][loop]$
            \State $\pi[i][(loop + 1)\ \bmod\ 2] \gets \pi[i][(loop + 1)\ \bmod\ 2] + dp[j] \alpha[i] \pi[j][loop]$
        \EndFor
    \EndFor
\EndWhile
\Statex
\Statex \textbf{Calculating the local fairness for each miner:}
\For{$i \in V$}
    \State $LF_1[i] \gets \pi[i][(loop + 1)\ \bmod\ 2] - \alpha[i]$
    \For{$j \in V$}
        \State $LF_1[i] \gets LF_1[i] + \pi[j] \alpha[i] F[j][i] (1 - W[j][i]) - \pi[i] \alpha[j] F[i][j] (1 - W[i][j])$
    \EndFor
    \State $LF_2[i] \gets LF_1[i] / \alpha[i]$ 
\EndFor

\end{algorithmic}
\end{algorithm}

\subsection{Parameters}\label{parameter}
\subsubsection{How to Determine $F_{ij}$}
Let $T$ denote the average block generation interval, and let $T_{ij}$ represent the time it takes for a block generated by miner $i$ to be received by miner $j$. Then, $F_{ij}$ is determined as follows:
\begin{align}
F_{ij} &= \int_{0}^{T_{ij}} \frac{e^{-\frac{x}{T}}}{T} \, dx \\
       &= 1 - e^{-\frac{T_{ij}}{T}}. \label{forkrateij}
\end{align}

\subsubsection{Tips for How to Determine $W_{ij}$}
Prior to any explanations regarding $W_{ij}$, first discussing the concept of chain ties is crucial.

Each miner constructs chains from their blocks and selects one main chain among them. The rule for selecting this chain is known as the fork choice rule. For instance, in Bitcoin, the longest chain rule, which selects the longest chain, is adopted.

However, in some cases, the fork choice rule alone may not uniquely determine the main chain owing to the occurrence of forks. This situation is known as a chain tie. A tie-breaking rule is implemented to resolve a chain tie. For instance, in Bitcoin, the first-seen rule, which selects the chain received first, is adopted.

We categorize practical tie-breaking rules as follows:
\begin{description}
  \item[\textbf{First-seen rule}] \mbox{}\\Selects the earliest arriving chain among the chains in a tie. Used in Bitcoin.
  \item[\textbf{Random rule}]  \mbox{}\\Randomly selects a chain among the chains in a tie~\cite{majorityisnotenough}. Proposed as a countermeasure to selfish mining. Used in Ethereum.
  \item[\textbf{Last-generated rule}] \mbox{}\\Selects the latest chain among the chains in a tie~\cite{oneweirdtricktostopselfishminers, sakurai2024tiebreaking, sakurai2024fullylocallastgeneratedrule}. Suppresses selfish mining more effectively than the random rule.
\end{description}

Next, we explain how to determine $W_{ij}$. The value of $W_{ij}$ is significantly influenced by the hashrate of miners mining on the block generated by miner $i$ during a fork. More specifically, $W_{ij}$ is largely affected by the following two factors:
\begin{description}
  \item[\textbf{Tie-breaking rule}] \mbox{}\\During a fork, chain ties often occur. The tie-breaking rule determines the block on which miners, other than the block generator, will mine.
  \item[\textbf{Proportion of hashrate of the block generator}] \mbox{}\\The block generator mines on its own generated block regardless of the tie-breaking rule.
\end{description}
Other factors, such as the block propagation time and the number of miners participating in the network, also influence $W_{ij}$. Section \ref{mbcvalidation} provides further details on specific methods for calculating $W_{ij}$.

\section{Validation} \label{mbcvalidation}
Herein, we validate the capability of our proposed model-based calculation to determine mining fairness accurately. First, we examine the assumption that the number of blocks per round is at most two from the perspective of the scale of the fork. Next, we compare the results of simulation experiments with those of the model-based calculation. 

While it is excessively time-consuming to calculate mining fairness via simulations of networks composed of many (approximately 100 or more) miners, it is feasible to calculate mining fairness accurately and relatively quickly for networks with a small number of miners (2--10). In this study, we perform the validation using networks comprising two and ten miners, demonstrating that the proposed model-based calculation determines mining fairness much more accurately than existing methods.

\subsection{Examining the Scale of Forks}\label{scale of fork}
The model-based calculation disregards the impact of large-scale forks. In particular, it assumes that the number of blocks per round is at most two. In this study, we investigate the effects of large-scale forks.

\textbf{Regarding the Scale of Forks: } First, we establish some facts regarding fork rates. Let the hashrate of miner $i$, where $i \in V$, be $M_i$. Let the total network hashrate be $M_{all}$. Additionally, let the probability of successfully generating a block with one hash calculation be $p$. Then, the average number of hash calculations required to generate a block is $1/p$. Therefore, the following equation holds:
\begin{align}
\frac{1}{p M_{all}} = T, \label{e1}
\end{align}
where $T$ is the average block generation interval. 

Next, let $N_i$ be the total number of hash calculations performed by miners who are unaware of the block generated by miner $i$. In this case, the following equation holds:
\begin{align}
N_i = \sum_{j \in V} M_j T_{ij}. \label{e2}
\end{align}

Let $T_{W, i}$ be the hashrate-weighted average block propagation time for the block generated by miner $i$. Then, the following equation holds:
\begin{align}
T_{W, i} = \sum_{j \in V} \alpha_j T_{ij}. \label{e3}
\end{align}

Therefore, from (\ref{e1}), (\ref{e2}), and (\ref{e3}), the following equation holds:
\begin{align}
p N_i &= p \sum_{j \in V} M_j T_{ij} \\
      &= \sum_{j \in V} \frac{M_j}{M_{all}} \frac{T_{ij}}{T} \\
      &= \frac{T_{W, i}}{T}. \label{ssf}
\end{align}

Next, we examine the occurrence rate of forks based on their scale. Let random variable $C_i$ denote the number of blocks in the round initiated by miner $i$. In this case, the following holds:
\begin{align}
P(C_i = 1) &= \sum_{j \in V} \alpha_j \int_{T_{ij}}^{\infty} e^{-\frac{x}{T}} dx \\
           &= \sum_{j \in V} \alpha_j e^{-\frac{T_{ij}}{T}}, \label{i1}
\end{align}
where $P(C_i = 1)$ denotes the probability that no forks occur. Then, the probability $P(C_i \neq 1)$ that a fork occurs is as follows:
\begin{align}
P(C_i \neq 1) &= 1 - P(C_i = 1) \\
              &= 1 - \sum_{j \in V} \alpha_j e^{-\frac{T_{ij}}{T}}. \label{forkratetw}
\end{align}

The probability that the number of blocks in a round will be three or more satisfies the following inequality:
\begin{align}
P(C_i \geq 3) &\leq \sum_{k=2}^\infty \binom{N_i}{k} p^k (1-p)^{N_i - k} \label{reduc} \\
              &= \sum_{k=2}^\infty \frac{N_i \cdots (N_i - k + 1)}{k!} p^k (1-p)^{N_i - k} \\
              &\leq \sum_{k=2}^\infty \frac{(p N_i)^k}{k!} e^{-p(N_i - k)} \\
              &= e^{-p N_i} \sum_{k=2}^\infty \frac{(e^p p N_i)^k}{k!} \\
              &= e^{-p N_i} (e^{e^p p N_i} - 1 - e^p p N_i) \label{ssb} \\
              &= e^{-\frac{T_{W, i}}{T}} (e^{e^p \frac{T_{W, i}}{T}} - 1 - e^p \frac{T_{W, i}}{T}) \label{ssa} \\
              &\xrightarrow[\frac{T_{W, i}}{T} \ \text{is constant}]{p \ \to \ 0} 1 - (1 + \frac{T_{W, i}}{T}) e^{- \frac{T_{W, i}}{T}}. \label{i3}
\end{align}
When the number of blocks in a round is two or more, at least three hash calculations succeed before all the blocks are fully shared, thus satisfying (\ref{reduc}). Equation (\ref{ssa}) is obtained by substituting (\ref{ssf}) into (\ref{ssb}).

From (\ref{i3}), it follows that the probability that the number of blocks in a round will be two satisfies the following inequality:
\begin{align}
P(C_i = 2) &= P(C_i \neq 1) - P(C_i \geq 3) \\
           &\geq \sum_{j \in V} \alpha_j (1 - e^{-\frac{T_{ij}}{T}}) - \left\{ 1 - (1 + \frac{T_{W, i}}{T}) e^{- \frac{T_{W, i}}{T}} \right\} \\
           &= (1 + \frac{T_{W, i}}{T}) e^{- \frac{T_{W, i}}{T}} - \sum_{j \in V} \alpha_j e^{- \frac{T_{ij}}{T}}. \label{i2}
\end{align}

\textbf{Impact By Fork Scale:} We define the impact $I_1$ for rounds with one block, impact $I_2$ for rounds with two blocks, and impact $I_3$ for rounds with three or more blocks as follows:
\begin{align}
I_1 &= e^{-\frac{d}{T}}, \\
I_2 &= (1 + \frac{d}{T}) e^{- \frac{d}{T}} - \sum_{j \in V} \alpha_j e^{- \frac{d}{T}} \\
    &= \frac{d}{T} e^{- \frac{d}{T}}, \\
I_3 &= 1 - (1 + \frac{d}{T}) e^{- \frac{d}{T}}.
\end{align}
These definitions are obtained by substituting $T_{ij} = d$ into (\ref{i1}), (\ref{reduc}), and (\ref{i3}). It should be noted that $I_2$ is defined based on the lower bound, whereas $I_3$ is defined based on the upper bound. In other words, $I_2$ is evaluated to be smaller, whereas $I_3$ is evaluated to be larger.

\begin{figure}[tb]
\begin{center}
\includegraphics[width=\linewidth]{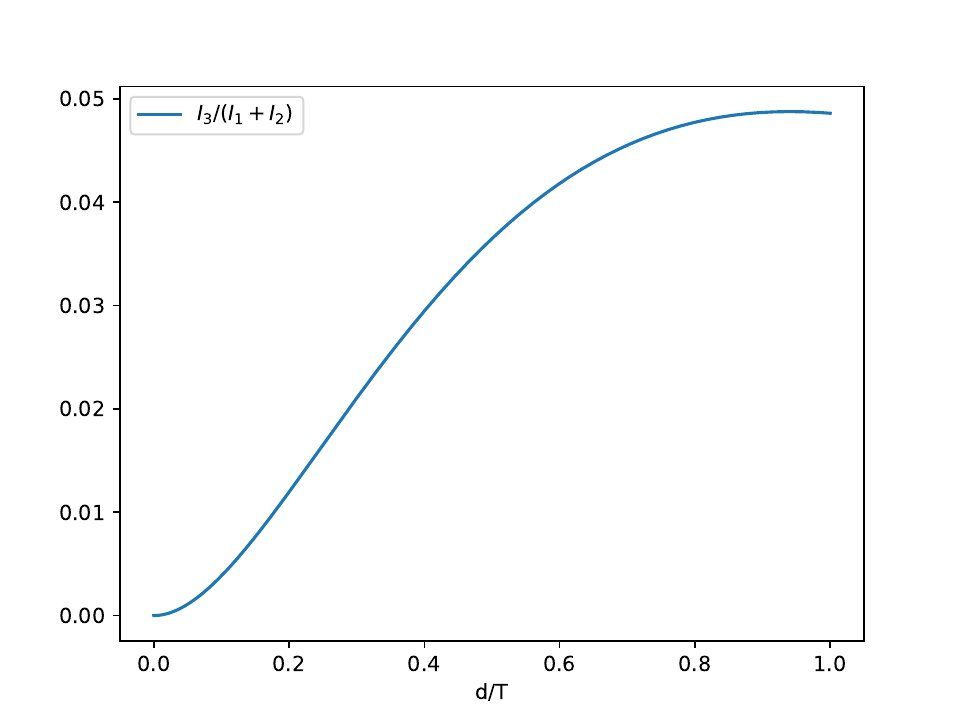}
\end{center}
\caption{Comparison between $I_3$ and $I_1+I_2$.}
\label{fig:I3_vs_I1+I2}
\end{figure}

\begin{figure}[tb]
\begin{center}
\includegraphics[width=\linewidth]{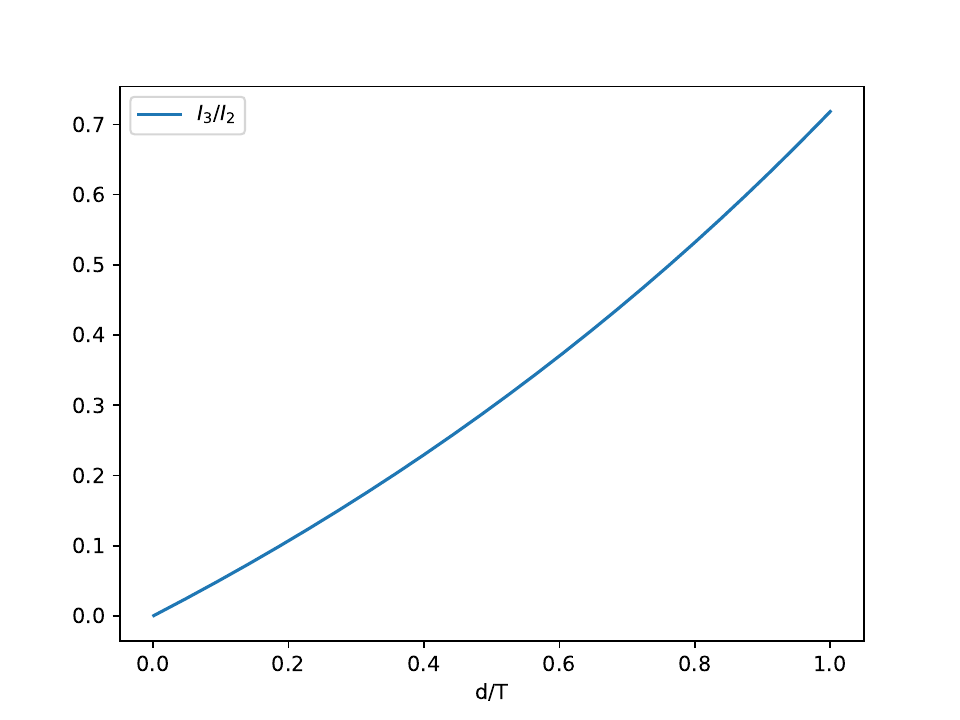}
\end{center}
\caption{Comparison between $I_3$ and $I_2$.}
\label{fig:I3_vs_I2}
\end{figure}

\begin{table}[tb]
\caption{Influence of $I_3$ Relative to $I_1$ and $I_2$.}
\label{ComI3I1I2}
\begin{center}
\begin{tabular}{|c|ccc|}
\hline
\multirow{2}{*}{}        & \multicolumn{3}{c|}{$d/T$}                             \\ \cline{2-4} 
                         & \multicolumn{1}{c|}{0.01} & \multicolumn{1}{c|}{0.1} & 0.5 \\ \hline
\multicolumn{1}{|c|}{$I_3/(I_1+I_2)$} & \multicolumn{1}{l|}{0.0000486868}    & \multicolumn{1}{l|}{0.00384871}   & \multicolumn{1}{l|}{0.0364743}   \\ \hline
\multicolumn{1}{|c|}{$I_3/I_2$} & \multicolumn{1}{l|}{0.0050167084}    & \multicolumn{1}{l|}{0.0517091}   & \multicolumn{1}{l|}{0.297442} \\ \hline
\end{tabular}
\end{center}
\end{table}

In the model-based calculation of mining fairness, we consider cases in which the number of blocks per round is two or fewer, thereby ignoring $I_3$. Thus, we compare $I_3$ with $I_1$ and $I_2$. Figs.~\ref{fig:I3_vs_I1+I2} and \ref{fig:I3_vs_I2} show $I_3/(I_1+I_2)$ and $I_3/I_2$, respectively, as $d/T$ is varied from $0$ to $1$. The specific numerical values are listed in Table \ref{ComI3I1I2}. It is evident that as $d/T$ decreases, the influence of $I_3$ diminishes. Although it is not directly demonstrated herein whether the model-based calculation can determine actual mining fairness, it can be inferred that the model-based calculation will be effective in scenarios where $d/T$ is small.

\subsection{Networks Composed of Two Miners} \label{2nodevalidation}
The comparison of the impact of forks by scale in Section \ref{scale of fork} offers intuitive insight into how the assumption of the model affects the model-based calculation. However, it does not address how accurately the model-based calculation matches the actual numerical results for mining fairness. In this section, we validate the model-based calculation using a simple network composed of two miners in blockchain network simulations.

\subsubsection{Model-Based Calculation}
We perform the model-based calculation for a network comprising two miners. The calculations follow the procedure outlined in Section \ref{calculating}.

Prior to these calculations, we first provide some relevant definitions. Let the two miners in the network be $Miner_A$ and $Miner_B$. Let the proportion of the hashrate of $Miner_A$ be $\alpha_A$ and that of $Miner_B$ be $\alpha_B$, where $\alpha_A + \alpha_B = 1$. Let $T$ be the average block generation interval, and $d$ be the block propagation time. Let $\pi_A$ and $\pi_B$ be the round start rates of $Miner_A$ and $Miner_B$, respectively. Next, we define $f$ as follows:
\begin{align}
f &= 1 - e^{-\frac{d}{T}},
\end{align}
where $f$ denotes the probability that $Miner_B$ (or $Miner_A$) will create a fork when $Miner_A$ (or $Miner_B$) starts a round and the other miner generates the next block.

Next, we calculate the round start rate. Considering that a sufficiently long time has passed, the following equation holds:
\begin{align}
& \pi_B = \pi_A(\alpha_B f\alpha_B + \alpha_B(1-f)) + \pi_B(\alpha_B + \alpha_Af\alpha_B) \\
\Rightarrow & \pi_B (1-\alpha_B - \alpha_A f \alpha_B)= \pi_A(\alpha_B f\alpha_B + \alpha_B(1-f)) \\
\Rightarrow & (1-\pi_A) (1-\alpha_B - \alpha_A f \alpha_B)= \pi_A(\alpha_B f\alpha_B + \alpha_B(1-f)) \\
\Rightarrow & \pi_A (f\alpha_B (\alpha_B - 1 - \alpha_A)+1)= 1-\alpha_B - \alpha_A f \alpha_B \\
\Rightarrow & \pi_A (1 - 2 \alpha_A \alpha_B f)= 1-\alpha_B - \alpha_A f \alpha_B \\
\Rightarrow & \pi_A = \alpha_A\frac{1 - \alpha_B f}{1 - 2 \alpha_A f \alpha_B}. 
\end{align}
Because $\pi_A$ has already been calculated, $\pi_B$ can then be determined as follows:
\begin{align}
\pi_B &= 1 - \pi_A \\
      &= \alpha_B\frac{1 - \alpha_A f}{1 - 2 \alpha_A f \alpha_B}.
\end{align}

Thereafter, we calculate the probability $W_{AB}$ that the block generated by $Miner_A$ is incorporated into the main chain if $Miner_B$ generates a block by forking immediately after $Miner_A$ starts a round. For simplicity, we assume that the block of $Miner_B$ conflicts with the chain of $Miner_A$; that is, we ignore the case in which the block height of $Miner_B$ is smaller than that of $Miner_A$.
\begin{align}
W_{AB} =& \overset{1}{\alpha_A} \overset{2}{\alpha_A} +  \overset{1}{\alpha_A} \overset{2}{\alpha_B} (1-f) \nonumber  \\
      &+ \overset{1}{\alpha_A} \overset{1}{\alpha_B} f \lbrace \overset{2}{\alpha_A} \overset{3}{\alpha_A} + \overset{2}{\alpha_A} \overset{3}{\alpha_B} (1-f) \nonumber\\ & + \overset{2}{\alpha_A} \overset{2}{\alpha_B} f(\cdots) + \overset{2}{\alpha_B} \overset{2}{\alpha_A} f (\cdots)\rbrace \nonumber \\
      &+ \overset{1}{\alpha_B} \overset{1}{\alpha_A} f \lbrace \overset{2}{\alpha_A} \overset{3}{\alpha_A} + \overset{2}{\alpha_A} \overset{3}{\alpha_B} (1-f) \nonumber\\ & + \overset{2}{\alpha_A} \overset{2}{\alpha_B} f(\cdots) + \overset{2}{\alpha_B} \overset{2}{\alpha_A} f (\cdots)\rbrace  \label{2winning}\\
      =& \alpha_A \alpha_A +  \alpha_A\alpha_B (1-f) \nonumber \\
      &+ 2\alpha_A\alpha_B f \lbrace \alpha_A\alpha_A +  \alpha_A\alpha_B (1-f) + 2\alpha_A\alpha_B f(\cdots) \rbrace \\
      =& \lbrace \alpha_A \alpha_A +  \alpha_A\alpha_B (1-f) \rbrace \frac{1}{1 - 2 \alpha_A \alpha_B f} \\
      =& \alpha_A\frac{1 - \alpha_B f}{1 - 2 \alpha_A \alpha_B f},
\end{align}
where $\alpha_A$ and $\alpha_B$ denote the probabilities of $Miner_A$ and $Miner_B$, respectively, generating a block. The superscript numbers on $\alpha_A$ and $\alpha_B$ indicate the differences in the block height from the block that initially caused the chain tie.
The value of $W_{BA}$ can also be derived from $W_{AB}$ as follows:
\begin{align}
W_{BA} &= \alpha_B\frac{1 - \alpha_A f}{1 - 2 \alpha_A \alpha_B f}.
\end{align}
Then, the following relationships hold:
\begin{align}
\pi_A &= W_{AB}, \\
\pi_B &= W_{BA}.
\end{align}

Using $\pi_A$ and $\pi_B$, we determine $LF_1(A)$ as follows:
\begin{align}
LF_1(A) = \pi_A + (\alpha_A - \alpha_B) f \pi_A \pi_B - \alpha_A.
\end{align}

\subsubsection{Simulation Settings} Based on the blockchain network simulator SimBlock~\cite{simblock}, we developed another event-driven simulator composed of two miners. Our simulator can simulate forks of any scale, similar to those in real blockchain systems.

We examined variations in $d/T$, i.e., the ratio of the block propagation time to the average block generation interval, with values of $0.1$, $0.3$, and $0.5$. The block propagation time was kept constant among all miners. 

We also examined variations in $\alpha_A$, i.e., the proportion of hashrate of miner $A$, with values of $0.1$ and $0.3$. A value of $0.5$ was not considered because, in this case, mining fairness is completely maintained because of the symmetry of the network.

Each simulation consisted of ten billion rounds.

\begin{table}[tb]
\caption{Errors Between Simulation and Model-Based Calculation Results of $LF_1$ for Network of Two Miners.}
\label{2noderelativeerror}
\begin{center}
\begin{tabular}{|cc|ccc|}
\hline
\multicolumn{2}{|c|}{\multirow{2}{*}{}} & \multicolumn{3}{c|}{$d/T$} \\ \cline{3-5} 
\multicolumn{2}{|c|}{} & \multicolumn{1}{c|}{0.1} & \multicolumn{1}{c|}{0.3} & \multicolumn{1}{c|}{0.5} \\ \hline
\multicolumn{1}{|c|}{\multirow{2}{*}{$\alpha_A$}} & \multicolumn{1}{c|}{0.3} & \multicolumn{1}{l|}{0.00059936} & \multicolumn{1}{l|}{0.00792584} & \multicolumn{1}{l|}{0.0203151} \\ \cline{2-5} 
\multicolumn{1}{|c|}{} & \multicolumn{1}{c|}{0.1} & \multicolumn{1}{l|}{0.000315594} & \multicolumn{1}{l|}{0.0031603} & \multicolumn{1}{l|}{0.00744887} \\ \hline
\end{tabular}
\end{center}
\end{table}

\subsubsection{Validation Results} 
The errors between the simulation and model-based calculation results are listed in Table \ref{2noderelativeerror}. Here, error is defined as the relative error as follows:
\begin{align}
  \frac{d_{euclid}(LF_{simulation}, LF_{MBC})}{d_{euclid}(LF_{simulation}, 0)}, \label{defrelativeerrord}
\end{align}
where $d$ is the Euclidean distance, $LF_{simulation}$ is the vector of the simulated values of local mining fairness for each miner, and $LF_{MBC}$ is the vector of the model-based calculated values of local mining fairness.

The error values indicate that the model-based calculation can compute mining fairness with high accuracy. Furthermore, it can be observed that the accuracy deteriorates as $d/T$ increases. This is likely because, as seen in Section \ref{scale of fork}, the impact of having more than three blocks per round becomes more significant as $d/T$ increases.

\subsection{Network Composed of Ten Miners} \label{multinodevalidation}
In this section, we validate the model-based calculation of mining fairness on a network comprising ten miners. Compared to a network with two miners, a network with ten miners introduces additional elements, including tie-breaking rules, hashrate distribution, and block propagation time; this allows us to demonstrate that the proposed model-based calculation method is effective even in more complex networks. Furthermore, we compare our method against a state-of-the-art approach \cite{ModelingtheImpactofNetwork}, highlighting its advantageous performance.

\subsubsection{$W_{ij}$ in a Network Composed of Multiple Miners}
In a network with more than two miners, it is necessary to consider tie-breaking rules. Here, we demonstrate how to determine $W_{ij}$ for a network with multiple miners according to different tie-breaking rules. We assume that all forks cause chain ties.

\textbf{First-Seen Rule:} We assume that miner $i$ starts a round, and then miner $j$ causes a chain tie in the same round. Let $p_{i, j, k}$ be the probability that miner $k$ mines on the block generated by miner $i$. The time $T_{ij}$ it takes for the block generated by miner $i$ to reach miner $j$ is assumed to be a fixed value that depends only on $i$ and $j$. 

When $T_{ik} < T_{jk}$, regardless of the time when miner $j$ generates the block, the block generated by miner $i$ will reach miner $k$ first, and hence, $p_{i, j, k} = 1$. Similarly, when $T_{ik} < T_{ij} + T_{jk}$, the block generated by miner $j$ will reach miner $k$ first, and hence, $p_{i, j, k} = 0$. In other cases, the following equation holds:
\begin{align}
p_{i, j, k} &= \frac{\int_{T_{ik}-T_{jk}}^{T_{ij}}\frac{e^{-\frac{x}{T}}}{T} dx}{F_{ij}} \label{choosei}\\
            &= \frac{e^{-\frac{T_{ik}- T_{jk}}{T}} - e^{-\frac{T_{ij}}{T}}}{1 - e^{-\frac{T_{ij}}{T}}}. \label{subFij}
\end{align}
Equation (\ref{choosei}) defines the probability that the block of miner $i$ reaches miner $k$ first under the condition that a chain tie occurs. Equation (\ref{subFij}) substitutes $F_{ij}$ into (\ref{choosei}) based on (\ref{forkrateij}).

From $p_{i, j, k}$, the value of $W_{ij}$ is determined as follows:
\begin{align}
W_{ij} &= \sum_{k \in V} \alpha_k p_{i, j, k}.
\end{align}

\textbf{Random Rule:} Herein, mining is performed by selecting a block randomly during a chain tie. The value of $W_{ij}$ is given by the following equation:
\begin{align}
W_{ij} &= \alpha_i + \frac{1 - \alpha_i - \alpha_j}{2}.
\end{align}

\textbf{Last-Generated Rule:} In this rule, mining is performed by selecting the most recently generated block during a chain tie. The value of $W_{ij}$ is given by the following equation:
\begin{align}
W_{ij} &= \alpha_i.
\end{align}

\begin{figure}[tb]
\begin{center}
\includegraphics[width= \linewidth]{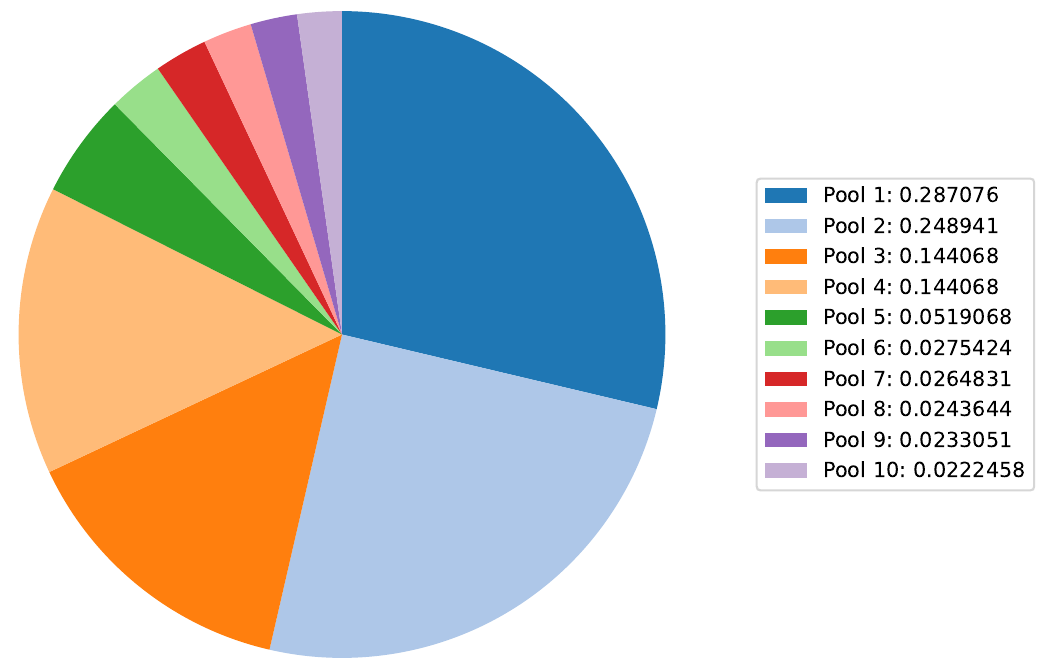}
\end{center}
\caption{Hashrate distribution settings used in validation of model-based calculation.}
\label{disuv}
\end{figure}

\begin{table*}[tb]
\caption{Mean and Standard Deviation (SD) of Errors Between Simulation and Model-Based Calculation Results for $d/T = 0.01$ and $0.04$ for Network of Ten Miners.}
\label{errormultip1}
\begin{center}
\begin{tabular}{|cc|cccccc|}
\hline
\multicolumn{2}{|c|}{\multirow{3}{*}{}}                  & \multicolumn{2}{c|}{First-seen rule}                  & \multicolumn{2}{c|}{Random rule}                      & \multicolumn{2}{c|}{Last-generated rule} \\ \cline{3-8} 
\multicolumn{2}{|c|}{}                                   & \multicolumn{6}{c|}{d/T}                                                                                                                                 \\ \cline{3-8} 
\multicolumn{2}{|c|}{}                                   & \multicolumn{1}{c|}{0.01} & \multicolumn{1}{c|}{0.04} & \multicolumn{1}{c|}{0.01} & \multicolumn{1}{c|}{0.04} & \multicolumn{1}{c|}{0.01}     & 0.04     \\ \hline
\multicolumn{1}{|c|}{\multirow{2}{*}{Round start rate}} & Mean & \multicolumn{1}{l|}{0.0000201254}    & \multicolumn{1}{l|}{0.000132747}   & \multicolumn{1}{l|}{0.0000208543}    & \multicolumn{1}{l|}{0.000134614}   & \multicolumn{1}{l|}{0.0000207662}        & \multicolumn{1}{l|}{0.000135}      \\ \cline{2-8} 
\multicolumn{1}{|c|}{}                            & SD   & \multicolumn{1}{l|}{0.00000639285}    & \multicolumn{1}{l|}{0.0000648318}   & \multicolumn{1}{l|}{0.00000663684}    & \multicolumn{1}{l|}{0.0000664632}   & \multicolumn{1}{l|}{0.00000700959}        & \multicolumn{1}{l|}{0.0000649111}      \\ \hline
\multicolumn{1}{|c|}{\multirow{2}{*}{$LF_1$}}      & Mean & \multicolumn{1}{l|}{0.00891706}    & \multicolumn{1}{l|}{0.0221225}    & \multicolumn{1}{l|}{0.0153555}    & \multicolumn{1}{l|}{0.0507862}    & \multicolumn{1}{l|}{0.0214307}        & \multicolumn{1}{l|}{0.080677}        \\ \cline{2-8} 
\multicolumn{1}{|c|}{}                            & SD   & \multicolumn{1}{l|}{0.00356624}    & \multicolumn{1}{l|}{0.0115312}    & \multicolumn{1}{l|}{0.00771275}    & \multicolumn{1}{l|}{0.0289467}    & \multicolumn{1}{l|}{0.0130309}        & \multicolumn{1}{l|}{0.0613974}       \\ \hline
\multicolumn{1}{|c|}{\multirow{2}{*}{$LF_2$}}      & Mean & \multicolumn{1}{l|}{0.0108707}    & \multicolumn{1}{l|}{0.0209106}    & \multicolumn{1}{l|}{0.0164265}    & \multicolumn{1}{l|}{0.0439469}    & \multicolumn{1}{l|}{0.0210495}        & \multicolumn{1}{l|}{0.0702858}        \\ \cline{2-8} 
\multicolumn{1}{|c|}{}                            & SD   & \multicolumn{1}{l|}{0.00373994}    & \multicolumn{1}{l|}{0.00828696}   & \multicolumn{1}{l|}{0.00546804}    & \multicolumn{1}{l|}{0.0146293}   & \multicolumn{1}{l|}{0.0110262}        & \multicolumn{1}{l|}{0.0426435}      \\ \hline
\end{tabular}
\end{center}
\end{table*}

\begin{table*}[tb]
\caption{Mean and Standard Deviation (SD) of Errors Between Simulation and Model-Based Calculation Results for $d/T = 0.07$ and $0.1$ for Network of Ten Miners.}
\label{errormultip2}
\begin{center}
\begin{tabular}{|cc|cccccc|}
\hline
\multicolumn{2}{|c|}{\multirow{3}{*}{}}                  & \multicolumn{2}{c|}{First-seen rule}                  & \multicolumn{2}{c|}{Random rule}                      & \multicolumn{2}{c|}{Last-generated rule} \\ \cline{3-8} 
\multicolumn{2}{|c|}{}                                   & \multicolumn{6}{c|}{d/T}                                                                                                                                 \\ \cline{3-8} 
\multicolumn{2}{|c|}{}                                   & \multicolumn{1}{c|}{0.07} & \multicolumn{1}{c|}{0.1} & \multicolumn{1}{c|}{0.07} & \multicolumn{1}{c|}{0.1} & \multicolumn{1}{c|}{0.07}     & 0.1     \\ \hline
\multicolumn{1}{|c|}{\multirow{2}{*}{Round start rate}} & Mean & \multicolumn{1}{l|}{0.000406944}    & \multicolumn{1}{l|}{0.0008385}   & \multicolumn{1}{l|}{0.0004072}    & \multicolumn{1}{l|}{0.000840025}   & \multicolumn{1}{l|}{0.000408239}        & \multicolumn{1}{l|}{0.000839217}       \\ \cline{2-8} 
\multicolumn{1}{|c|}{}                            & SD   & \multicolumn{1}{l|}{0.000215053}    & \multicolumn{1}{l|}{0.000462696}   & \multicolumn{1}{l|}{0.000215627}    & \multicolumn{1}{l|}{0.000461556}   & \multicolumn{1}{l|}{0.000217068}        & \multicolumn{1}{l|}{0.000460759}      \\ \hline
\multicolumn{1}{|c|}{\multirow{2}{*}{$LF_1$}}      & Mean & \multicolumn{1}{l|}{0.0851189}    & \multicolumn{1}{l|}{0.0553571}    & \multicolumn{1}{l|}{0.0851189}    & \multicolumn{1}{l|}{0.117411}    & \multicolumn{1}{l|}{0.13679}        & \multicolumn{1}{l|}{0.188992}       \\ \cline{2-8} 
\multicolumn{1}{|c|}{}                            & SD   & \multicolumn{1}{l|}{0.0467998}    & \multicolumn{1}{l|}{0.0293375}    & \multicolumn{1}{l|}{0.0467998}    & \multicolumn{1}{l|}{0.0630781}    & \multicolumn{1}{l|}{0.104206}        & \multicolumn{1}{l|}{0.141844}       \\ \hline
\multicolumn{1}{|c|}{\multirow{2}{*}{$LF_2$}}      & Mean & \multicolumn{1}{l|}{0.0737303}    & \multicolumn{1}{l|}{0.0520755,}    & \multicolumn{1}{l|}{0.0737303}    & \multicolumn{1}{l|}{0.10182}    & \multicolumn{1}{l|}{0.117051}        & \multicolumn{1}{l|}{0.159697}      \\ \cline{2-8} 
\multicolumn{1}{|c|}{}                            & SD   & \multicolumn{1}{l|}{0.0242187}    & \multicolumn{1}{l|}{0.0205341}   & \multicolumn{1}{l|}{0.0242187}    & \multicolumn{1}{l|}{0.0328015}   & \multicolumn{1}{l|}{0.0670573}        & \multicolumn{1}{l|}{0.0871817}    \\ \hline
\end{tabular}
\end{center}
\end{table*}

\begin{table*}[tb]
\caption{Proposed Model-Based Calculation of $LF_1$ for Network of Ten Miners vs.~Existing Method.}
\label{multinodecomparlf1eerror}
\begin{center}
\begin{tabular}{|cc|cccc|}
\hline
\multicolumn{2}{|c|}{\multirow{2}{*}{}} & \multicolumn{4}{c|}{$d/T$} \\ \cline{3-6} 
\multicolumn{2}{|c|}{} & \multicolumn{1}{c|}{0.01} & \multicolumn{1}{c|}{0.04} & \multicolumn{1}{c|}{0.07} & \multicolumn{1}{c|}{0.1} \\ \hline
\multicolumn{1}{|c|}{\multirow{2}{*}{Fist-seen rule}} & \multicolumn{1}{c|}{Proposed method} & \multicolumn{1}{l|}{0.00891706} & \multicolumn{1}{l|}{0.0221225} & \multicolumn{1}{l|}{0.0386387} & \multicolumn{1}{l|}{0.0553571} \\ \cline{2-6} 
\multicolumn{1}{|c|}{} & \multicolumn{1}{c|}{Existing method} & \multicolumn{1}{l|}{1.23734} & \multicolumn{1}{l|}{1.23344} & \multicolumn{1}{l|}{1.22872} & \multicolumn{1}{l|}{1.22402}  \\ \hline
\multicolumn{1}{|c|}{\multirow{2}{*}{Random rule}} & \multicolumn{1}{c|}{Proposed method} & \multicolumn{1}{l|}{0.0153555} & \multicolumn{1}{l|}{0.0507862} & \multicolumn{1}{l|}{0.0851189} & \multicolumn{1}{l|}{0.117411} \\ \cline{2-6} 
\multicolumn{1}{|c|}{} & \multicolumn{1}{c|}{Existing method} & \multicolumn{1}{l|}{1.69555} & \multicolumn{1}{l|}{1.66842} & \multicolumn{1}{l|}{1.64136} & \multicolumn{1}{l|}{1.61529}  \\ \hline
\multicolumn{1}{|c|}{\multirow{2}{*}{Last-generated rule}} & \multicolumn{1}{c|}{Proposed method} & \multicolumn{1}{l|}{0.0214307} & \multicolumn{1}{l|}{0.080677} & \multicolumn{1}{l|}{0.13679} & \multicolumn{1}{l|}{0.188992} \\ \cline{2-6} 
\multicolumn{1}{|c|}{} & \multicolumn{1}{c|}{Existing method} & \multicolumn{1}{l|}{1.56918} & \multicolumn{1}{l|}{1.59359} & \multicolumn{1}{l|}{1.60905} & \multicolumn{1}{l|}{1.6178}  \\ \hline
\end{tabular}
\end{center}
\end{table*}

\begin{table*}[tb]
\caption{Proposed Model-Based Calculation of $LF_2$ for Network of Ten Miners vs.~Existing Method.}
\label{multinodecomparlf2eerror}
\begin{center}
\begin{tabular}{|cc|cccc|}
\hline
\multicolumn{2}{|c|}{\multirow{2}{*}{}} & \multicolumn{4}{c|}{$d/T$} \\ \cline{3-6} 
\multicolumn{2}{|c|}{} & \multicolumn{1}{c|}{0.01} & \multicolumn{1}{c|}{0.04} & \multicolumn{1}{c|}{0.07} & \multicolumn{1}{c|}{0.1} \\ \hline
\multicolumn{1}{|c|}{\multirow{2}{*}{Fist-seen rule}} & \multicolumn{1}{c|}{Proposed method} & \multicolumn{1}{l|}{0.0108707} & \multicolumn{1}{l|}{0.0209106} & \multicolumn{1}{l|}{0.0365199} & \multicolumn{1}{l|}{0.0520755} \\ \cline{2-6} 
\multicolumn{1}{|c|}{} & \multicolumn{1}{c|}{Existing method} & \multicolumn{1}{l|}{0.913274} & \multicolumn{1}{l|}{0.910973} & \multicolumn{1}{l|}{0.908631} & \multicolumn{1}{l|}{0.906371}  \\ \hline
\multicolumn{1}{|c|}{\multirow{2}{*}{Random rule}} & \multicolumn{1}{c|}{Proposed method} & \multicolumn{1}{l|}{0.0164265} & \multicolumn{1}{l|}{0.0439469} & \multicolumn{1}{l|}{0.0737303} & \multicolumn{1}{l|}{0.10182} \\ \cline{2-6} 
\multicolumn{1}{|c|}{} & \multicolumn{1}{c|}{Existing method} & \multicolumn{1}{l|}{1.11546} & \multicolumn{1}{l|}{1.10628} & \multicolumn{1}{l|}{1.09711} & \multicolumn{1}{l|}{1.08832}  \\ \hline
\multicolumn{1}{|c|}{\multirow{2}{*}{Last-generated rule}} & \multicolumn{1}{c|}{Proposed method} & \multicolumn{1}{l|}{0.0210495} & \multicolumn{1}{l|}{0.0702858} & \multicolumn{1}{l|}{0.117051} & \multicolumn{1}{l|}{0.159697} \\ \cline{2-6} 
\multicolumn{1}{|c|}{} & \multicolumn{1}{c|}{Existing method} & \multicolumn{1}{l|}{1.1644} & \multicolumn{1}{l|}{1.16133} & \multicolumn{1}{l|}{1.15597} & \multicolumn{1}{l|}{1.15099}  \\ \hline
\end{tabular}
\end{center}
\end{table*}

\subsubsection{Simulation Settings}

The simulator used in this validation was an extended version of a network simulator composed of two miners. The number of miners was set to ten. The hashrate distribution was based on that of Bitcoin~\cite{MiningPoolStats}. The hashrate distribution settings are illustrated in Fig.~\ref{disuv}.

In this validation, the ratios of the average block propagation time to the average block generation interval, $d/T$, were varied among the values of $0.01$, $0.04$, $0.07$, and $0.1$. These settings cover most blockchain systems; for instance, in the case of Bitcoin, $d/T$ is approximately equal to $0.00576$, whereas in the case of Ethereum, $d/T = 0.068$~\cite{onTheSecurity, Calibratingtheperformance}. The block propagation time distribution among the different miners followed an exponential distribution~\cite{informationpropagation}, whereas the block propagation time to oneself was set to 0. Additionally, the previously described tie-breaking rules, i.e., the first-seen rule, random rule, and last-generated rule, were examined. 

Each simulation consisted of ten billion rounds.

\subsubsection{Validation Results} 
We conducted 50 simulation experiments for each validation target.
The errors between the simulation and model-based calculation results are listed in Tables \ref{errormultip1} and \ref{errormultip2}. Here, error is defined as the relative error as in (\ref{defrelativeerrord}). The difference from the previous validation is that the number of elements in the $LF$ vector is changed from two to ten, and we investigate not only the $LF$ vector but also the round-start-rate vector.

First, we examine the round start rate. As observed, the model-based calculations match the simulation results with high accuracy under all conditions. 

Next, we examine mining fairness. The tables demonstrate that the model-based calculation method can compute mining fairness with high accuracy; additionally, it can be observed that the accuracy deteriorates as $d/T$ increases.

Furthermore, it is observed that the accuracy of the mining fairness calculations is not as high as that of the round-start-rate calculations or of the mining fairness calculations for a network composed of two miners; this is because the mining fairness calculation requires $W_{ij}$.

Additionally, it is observed that the first-seen rule is more accurate than the random rule or the last-generated rule. This finding indicates that the calculation of $W_{ij}$ based on the first-seen rule is superior; this is because the influence of blocks up to the second one in a given round is stronger under the first-seen rule.

We also compared our proposed model-based calculation with a state-of-the-art method. This comparison method is the same as that proposed except that the latter considers the round start rate \cite{ModelingtheImpactofNetwork}. Conversely, the comparison method automatically assumes that the round start rate is equal to the proportion of hashrate. Tables~\ref{multinodecomparlf1eerror} and \ref{multinodecomparlf2eerror} present the results. As demonstrated, our model-based calculation significantly improves the accuracy compared with that of the state-of-the-art method. This result demonstrates the importance of considering the impact of forks on the round start rate.

\section{Conclusion}\label{conclusion}
In this paper, we propose an efficient method for calculating mining fairness, one of the key metrics in blockchain, by approximating a complex blockchain network with a simpler network, where the number of blocks per round is at most two. Through simulation experiments, we demonstrated that our approach significantly enhances the accuracy of mining fairness calculations compared to existing methods. We anticipate that our contributions will stimulate further research on mining fairness across various domains, including block propagation protocols, neighbor node selection methods, and pool-selection strategies.


\printbibliography

\end{document}